\documentclass[aps,prl,twocolumn,floatfix,superscriptaddress,showpacs,amsfonts
,amssymb,amsmath,preprintnumbers]{revtex4-1}
\usepackage{bm}
\usepackage{epsfig}
\usepackage{graphicx}
\usepackage{color}
\usepackage{subfigure}

\newcommand{\eq}[1]{Eq.~(\ref{#1})}
\newcommand{\eqs}[2]{Eqs.~(\ref{#1}--\ref{#2})}

\newcommand{\be}{\begin{equation}}
\newcommand{\ee}{\end{equation}}
\newcommand{\beq}{\begin{equation}}
\newcommand{\eeq}{\end{equation}}

\newcommand\bea{\begin{eqnarray}}
\newcommand\eea{\end{eqnarray}}

\newcommand{\Dprime}{\Delta'}
\newcommand{\aX}{a_{X,N}}

\newcommand{\vdr}{v_{\rm dr}}
\newcommand{\tdr}{\tau_{\rm dr}}
\newcommand{\Mdr}{M_{\rm dr}}
\newcommand{\Mdrc}{M_{\rm dr,c}}

\newcommand{\tauAck}{\tau_{A0}}

\newcommand{\tcr}{t_{\rm cr}}
\newcommand{\tcrFKR}{t_{\rm cr}^{\rm FKR}}
\newcommand{\tcrCoppi}{t_{\rm cr}^{\rm Coppi}}

\newcommand{\ttr}{t_{\rm tr}}

\newcommand{\NmaxCoppi}{N_{\rm max}^{\rm Coppi}}
\newcommand{\kmaxCoppi}{k_{\rm max}^{\rm Coppi}}
\newcommand{\gmaxCoppi}{\gamma_{\rm max}^{\rm Coppi}}

\newcommand{\gmaxFKR}{\gamma_{\rm max}^{\rm FKR}}

\newcommand{\tdisruptCoppi}{t_{\rm disrupt}^{\rm Coppi}}
\newcommand{\tdisruptFKR}{t_{\rm disrupt}^{\rm FKR}}

\newcommand{\tdisrupt}{t_{\rm disrupt}}
\newcommand{\deltain}{\delta_{\rm in}}

\renewcommand{\[}{\left[}
\renewcommand{\]}{\right]}

\begin{document}

\title{Magnetic Reconnection Onset via Disruption of a Forming 
Current Sheet\\ by the Tearing Instability}
\author{D.\ A.\ Uzdensky}
\affiliation{Center for Integrated Plasma Studies, Physics Department, 
UCB-390, University of Colorado, Boulder Colorado 80309, USA}
\author{N.\ F.\ Loureiro}
\affiliation{Instituto de Plasmas e Fus\~ao Nuclear,
Instituto Superior T\'ecnico, \\
Universidade de Lisboa, 1049-001 Lisboa, Portugal}
\affiliation{Plasma Science and Fusion Center, Massachusetts Institute of Technology, Cambridge MA 02139, USA}

\date{\today}

\begin{abstract}
The recent realization that Sweet-Parker current sheets
 are violently unstable to the secondary tearing (plasmoid) instability 
implies that such current sheets cannot occur in real systems. 
This suggests that, in order to understand the onset of magnetic reconnection, 
one needs to consider the growth of the tearing instability in a current layer 
as it is being formed. 
Such an analysis is performed here in 
the context of nonlinear resistive MHD for a generic time-dependent 
equilibrium representing a gradually forming current sheet. 
It is shown that two onset regimes, single-island and multi-island, are possible, 
depending on the rate of current sheet formation. 
A simple model is used to compute the criterion for transition between these two 
regimes, as well as the reconnection onset time and the current sheet parameters at 
that moment. For typical solar corona parameters this model yields results
consistent with observations.

\end{abstract}

\pacs{52.35.Vd, 96.60.Iv, 52.30.Cv}

\maketitle


\paragraph{Introduction.}
Magnetic reconnection is a basic plasma process responsible for solar flares, magnetospheric substorms, and tokamak disruptions 
~\cite{biskamp_magnetic_2005,zweibel_magnetic_2009,yamada_magnetic_2010}.
While reconnection itself has been intensely studied, 
{\it reconnection onset} --- the transition from a slow quiescent stage of magnetic energy accumulation to an explosive energy release  --- is much less understood 
and remains one of the most mysterious aspects of this fascinating  phenomenon
~\cite{moore_triggering_1992,bhattacharjee_impulsive_2004,chifor_X-ray_2007, katz_laboratory_2010, joshi_pre-flare_2011,shibata_solar_2011}.

Reconnection is associated with quasi-two-dimensional intense electric current sheets (CSs) that can form in a plasma.  Although several special cases of CS formation have been investigated 
~\cite{chapman_liquid_1963,syrovatskii_formation_1971, hahm_forced_1985, 
waelbroeck_current_1989, wang_forced_1992, waelbroeck_onset_1993, longcope_current_1996,
priest_magnetic_2002, sui_evidence_2003, uzdensky_partial_2002, loureiro_x-point_2005}, 
a solid, general understanding of CS formation is still lacking.
Consequently, most numerical studies of reconnection are initialized with a fully developed CS, e.g., a resistive Sweet-Parker (SP) layer~\cite{parker_sweet_1957,sweet_neutral_1958}.  
However, the recent realization that long SP-like CSs are super-Alfv\'enically unstable implies that, in reality, they can never form in the first place~\cite{loureiro_instability_2007,lapenta_self_2008,bhattacharjee_fast_2009,samtaney_formation_2009, loureiro_plasmoid_2013,pucci_reconnection_2014,loureiro_magnetic_2016}.
This is also true for collisionless systems~\cite{daughton_fully_2006}.
Thus, reconnection onset needs to be investigated in the context of a gradual CS formation process, and 
addressing this important problem is the main goal of this Letter.


\paragraph{Problem setup.}
We consider a CS whose key parameters (thickness~$a$, length~$L$, and the reversing magnetic field~$B_0$) are slowly evolving on some time scale~$\tdr$, with the aspect ratio $L/a$ increasing.
With time, the system becomes unstable to multiple tearing modes, each characterized by a wavenumber~$k(t)$, the number of islands $N \sim k L$, and an amplitude (island width) $w_N(t)$.
Our goal is to analyze both the linear and nonlinear evolution of these modes in a forming CS, and to identify the first mode that exceeds the CS width~$a$, thus effectively disrupting the forming CS and 
marking the transition from the slow energy build-up stage to reconnection onset~\footnote{While the onset of reconnection involves a single dominant (primary) mode, a broad power-law distribution of island sizes may form during the following main reconnection stage~\cite{uzdensky_fast_2010, fermo_statistical_2010, loureiro_magnetic_2012, huang_distribution_2012}}. 

Whereas here we present only a resistive MHD analysis, the underlying conceptual framework should also be valid for weakly-collisional plasmas. 
Also, we ignore the effects of background sheared flows (e.g., associated with CS formation) on tearing evolution~\cite{bulanov_tearing_1979}, which we have found to be justified for sub-Alfv\'enic flows~\cite{loureiro_flows_2016}.


\paragraph{Linear stage.}
Tearing modes are linearly unstable if the instability parameter $\Delta'(k)>0$~\cite{FKR}.
For definiteness, consider a Harris-type magnetic equilibrium~\cite{Harris_1962}, for which $\Delta'a=2(1/ka-ka)$, and focus on long wavelength modes, $ka \ll 1$, so that $\Delta' a \sim 1/ka $. 
There are two possible linear regimes: (1) $\Delta'\deltain \ll 1$ (`FKR'~\cite{FKR}); 
and (2) $\Delta'\deltain \sim 1$  (`Coppi'~\cite{coppi_resistive_1976}). 
Here $\deltain = \[\gamma (k V_A)^{-2}a^2\eta\]^{1/4}$ is the inner resistive layer width, 
$\gamma$ is the growth rate, $V_A$ is the Alfv\'en speed, and $\eta$ is magnetic diffusivity. 
In the FKR case, $\gamma^{\rm FKR} \simeq \Delta'^{4/5}k^{2/5}V_A^{2/5}a^{-2/5}\eta^{3/5} \propto k^{-2/5}$, and so the fastest growing FKR mode is the longest that fits in the CS, $N \sim kL\sim 1$, corresponding to $\gmaxFKR \simeq L^{2/5}V_A^{2/5}a^{-2}\eta^{3/5}= 
\tau_A^{-1}S_a^{-3/5}(L/a)^{2/5}$, where $S_a \equiv a V_A/\eta \gg 1$ and $\tau_A \equiv a/V_A$.

In the Coppi case, $\gamma^{\rm Coppi} \simeq k^{2/3}V_A^{2/3}a^{-2/3}\eta^{1/3}=
\tau_A^{-1}S_a^{-1/3}N^{2/3}(L/a)^{-2/3} \propto k^{2/3}$.
The fact that the FKR and Coppi scalings of $\gamma$ with $k$ have different signs implies that the overall fastest mode is found by balancing the two expressions for~$\gamma$. 
This ``transitional" or ``fastest Coppi" mode has $\kmaxCoppi a\sim S_a^{-1/4}$ and $\gmaxCoppi \sim \tau_A^{-1}\, S_a^{-1/2}$. 
Modes with $k> \kmaxCoppi$ are in the FKR regime, while those with $k< \kmaxCoppi$ are in the Coppi regime; in a given CS, Coppi modes exist only if the $\kmaxCoppi$ mode fits inside it, $\kmaxCoppi L > 1$.


As the sheet's aspect ratio $L/a$ increases over time, higher and higher-$N$ modes become progressively  destabilized. Thus, many tearing modes may independently undergo linear evolution; their amplitudes during the linear stage are smaller than~$a$, so they do not yet affect CS formation or each other.  
For any given mode~$N$ we identify two important moments, marking transitions between different stages in the mode's life: the time $\tcr(N)$ marking the end of the linear stage, and the time $\ttr(N)$ for the mode's transition from the FKR to the Coppi regime.  The life path of the mode then depends on the relative ordering of $\tcr(N)$ and~$\ttr(N)$.  We define these two times as follows. 

First, both for FKR modes (for any given $N$) and for the fastest Coppi mode [$N(t) = N_{\rm max}^{\rm Coppi} \equiv k_{\rm max}^{\rm Coppi}(t) L(t)$], the linear growth rate $\gamma(N,t)$ increases with time as the CS develops ($L/a$ grows).
At some critical time~$\tcr(N)$, $\gamma(N,t)$ overcomes driving, $\gamma(N,\tcr) \gtrsim \tdr^{-1}$. 
After $\tcr(N)$ is reached and until the end of the linear stage, the mode's growth effectively proceeds on a frozen background and so, ignoring logarithmic corrections, $\tcr(N)$ marks the end of the mode's linear stage.

Second, whereas each mode $N$ always starts out in the FKR regime ($\NmaxCoppi(t) < N$), as the CS forms and hence both $L/a$ and $\NmaxCoppi \sim (L/a) S_a^{-1/4}$ increase, there may come a time $\ttr(N)$ when $\NmaxCoppi[\ttr(N)]=N$. 
At this point, if the mode is still in the linear regime, i.e., if $\ttr(N) < \tcr(N)$, the mode transitions to the Coppi regime. Importantly, higher~$N$ corresponds to higher~$\ttr(N)$.

Our main goal in the linear evolution analysis is to find which mode has the earliest~$\tcr(N)$ and hence reaches the end of its linear stage first. 
At early times, the fastest-growing mode is always the $N=1$ ($kL \sim 1$) FKR mode but, in principle, this may change over time. 
If $\tcr(1)\equiv \tcr(N=1)< \ttr(1)$ (we call this case the ``FKR scenario"), then the $N=1$ mode remains in the FKR regime throughout its linear evolution.  
Furthermore, since $\ttr(N)$ increases with~$N$, $\ttr(N)> \tcr(1)$, and so all the higher-$N$ modes also remain in the FKR regime throughout this time and thus grow slower than the $N=1$ mode. 
Thus, the $N=1$ FKR mode reaches its $\tcr$ first and thus ``wins" the linear stage. 

If, however, $\tcr(1)>\ttr(1)$, and so $L/a > S_a^{1/4}$ before $\tcr(1)$ is reached, then the $N=1$ mode transitions to the Coppi regime.   After that, the fastest growing linear mode at any given time is the transitional mode $\NmaxCoppi(t) = (L/a)\, S_a^{-1/4}> 1$. 
Then, the mode that actually wins the linear stage is the one with $\ttr(N) = \tcr(N)$, i.e., the mode that reaches its $\tcr$ immediately upon transitioning to the Coppi regime and hence becoming fastest-growing --- before it is overtaken by another Coppi mode.  

At the end of the linear stage, the amplitude of any mode is $w_N(\tcr)\sim\delta_{in}(k)\ll a(\tcr)$, and thus, 
to determine when the CS is disrupted ($w\sim a$) and by which mode, we now need to consider the nonlinear evolution.
 

\paragraph{Nonlinear Stage.}
The nonlinear evolution of a given mode $N$ is governed by the product~$\Delta'(k_N) w_N$.
If it is small at the nonlinear phase onset, $t=\tcr(N)$, then the mode enters the algebraic-growth Rutherford stage~\cite{rutherford_nonlinear_1973}
\footnote{Whereas Rutherford~\cite{rutherford_nonlinear_1973} assumes that only the $N=1$ harmonic is unstable but all other modes are stable, his results actually remain valid even when $N>1$ modes are also unstable~\cite{arcis_influence_2009}, as we assume here.}, 
 $\dot{w}_N  \simeq \eta \, \Delta'(k_N,t)$.
This is the case for modes that are in the FKR regime at~$\tcr(N)$. 
Indeed, the condition for $\Delta'(N) w_N \sim \deltain\Delta' \simeq \[L/aN\]^{8/5} S_a^{-2/5}$ at $t=\tcr(N)$ to be small is equivalent to the condition, $L(\tcr)/a(\tcr) < N S_a^{1/4}$, for them to be in the FKR regime in the first place. Thus, these modes have a well-defined nonlinear Rutherford stage that lasts until $w_N \sim 1/\Delta'$.

However, as both the island width $w_N$ and the CS aspect ratio grow, $\Delta' w_N \sim (w_N/a)\, (L/Na)$ also grows. Eventually, at some $t=t_X(N)$, the mode reaches the critical amplitude $w_{X,N} \sim 1/\Delta'(k_N)$ for undergoing $X$-point collapse~\cite{waelbroeck_onset_1993,loureiro_x-point_2005, militello_deceiving_2014}.  The Rutherford stage then ends, the inter-island X-points rapidly collapse to thin secondary current sheets, and the mode's growth greatly accelerates~\cite{loureiro_x-point_2005}. 
Since $w_{X,N} \sim [\Delta'(k_N)]^{-1} \sim k_N a^2  \ll a$, the dominant tearing mode has to undergo X-point collapse before it can disrupt the sheet ($w_{{\rm disrupt},N} = a$). 
However, because of the rapid, exponential post-collapse growth, the delay between these two events is short,  and so the first mode to reach $X$-point collapse remains dominant throughout the subsequent post-collapse  evolution and thus becomes the one that disrupts the~CS.  
Our goal is to identify this dominant mode and determine its collapse time, $t_{X,N}$, which then gives a good estimate for the CS disruption time, $\tdisrupt$.

First, consider the case $\tcr(1) < \ttr(1)$, when the first mode to end its linear stage is the FKR $N=1$ mode. Then, a comparison of the $N$-dependencies of $\tcr(N)$ and~$\ttr(N)$ shows that $\tcr(N)< \ttr(N)$ also for all other modes and thus they all remain in the FKR regime throughout linear evolution and then transition to the Rutherford stage. There are no Coppi modes in this case. 

Furthermore, from the Rutherford growth equation, $\dot{w}_N \simeq \eta \Delta'(k_N) \sim k^{-1} \sim N^{-1}$, and so lower-$N$ modes grow faster than higher-$N$ modes during this stage, implying that 
the $N=1$ mode will reach $X$-point collapse first.
More rigorously, integrating $\dot{w}_N \simeq \eta \Delta'(k_N)$ yields
\beq
\label{eq_Ruth_w_t}
w_N(t) = w_N(\tcr) + {2\eta\over N}\, \int_{\tcr}^t \, {{L(t')}\over{a^2(t')}} \, dt' \, ,
\eeq
where $w_N(\tcr)\sim \delta_{\rm in}\ll a(\tcr)$.

For simplicity, let us first assume that $a$ decreases with time, $\dot{a} < 0$ 
($L$ and $B_0$ may also be changing); we can then parameterize the CS evolution by $a$ instead of~$t$. 
\eq{eq_Ruth_w_t} then yields  
\beq
\label{eq_wn_a}
w_N(a) \sim  {\eta\over N}\, \int_{a[\tcr(N)]}^a \, {{L(a)}\over{a^2}} \, 
t'(a) \, da \, ,
\eeq
where $t'(a) \equiv dt(a)/da < 0$ and where we have ignored~$w_N(\tcr)$ and factors of~$2$.
Then, the value $a=\aX$ at which $w_N(a)\Delta'\sim 1$, is implicitly given by 
\beq
N= \eta \, \frac{L(\aX)}{\aX^2} \, \int_{a[\tcr(N)]}^{\aX} \, 
{{L(a)}\over{a^2}} \, t'(a) \, da \, .
\eeq

Since the nonlinear Rutherford stage is generally much longer than the linear one (because the Rutherford growth is algebraic in time, whereas linear-stage growth is exponential), 
we can neglect the lower bound in this integral. 
Then, since $t'(a) < 0$, we see that the integral on the right-hand side (RHS), and thus the entire RHS,   decrease with~$\aX$.
Correspondingly, smaller-$N$ Rutherford modes have larger $\aX$ and thus reach the $X$-point collapse sooner. 
Moreover, one can show that this is also true for a CS that forms not by thinning ($\dot{a}<0$), but instead by stretching ($\dot{L}>0$) of the layer at a fixed thickness~$a_0$, as is the case for channel flows driven by the magnetorotational instability~\cite{goodman_parasitic_1994, pessah_angular_2010} in accretion disks.

To sum up, the nonlinear Rutherford evolution of FKR-type modes does not change the result of the linear analysis: if the most unstable linear mode was in the FKR regime ($N=1$ FKR mode), then this mode will continue to dominate during the Rutherford stage and will reach $X$-point collapse first. Ultimately, this is the mode that disrupts the sheet.

Now we consider the nonlinear evolution for the second case, $\tcr(1) > \ttr(1)$, when the first mode to reach the end of the linear stage is the fastest Coppi mode $N=\NmaxCoppi$. 
One can show that $\deltain\Delta'\sim 1$ at $t=\tcr$ for this mode, and hence the Rutherford regime is essentially absent and $X$-point collapse occurs promptly, soon upon the mode's arrival at the nonlinear stage. 
Correspondingly, this mode is the first to undergo X-point collapse and to ultimately disrupt the~CS; the disruption time is then comparable to the time spent in the linear regime, 
$\tdisruptCoppi\sim\tcr(\NmaxCoppi)$.

The main general conclusion from these considerations is that the outcome of the nonlinear evolution of both FKR/Rutherford and Coppi modes in a gradually forming CS is the same as in the linear stage: 
the first mode to reach the end of its linear stage ($\gamma \tau_{\rm dr} = 1$) will also be the first to undergo  $X$-point collapse ($w\Delta' \sim 1$) and subsequently will disrupt the CS ($w=a$). 


The above formalism is general and can be applied to any CS formation process if the functional forms for the time evolution of the sheet parameters are known.  A simple but general example is analyzed next.


\paragraph{Example: Chapman-Kendall-like current sheet formation.}
Consider an $X$-point configuration given by $\phi = \vdr x y /L(t)$, 
$\psi = B_0/2 [x^2/a(t)-y^2/L(t)]$, where $\phi$ is the (incompressible) flow stream function, 
$\psi$ is the magnetic flux, $B_0=\text{const}$, and $\vdr=\text{const}$ 
is the plasma velocity driving the CS formation (cf.~\cite{chapman_liquid_1963};~\footnote{The original 
Chapman-Kendall~\cite{chapman_liquid_1963} solution specifies 
$\phi=\Lambda(t) x y$, which, upon setting $\Lambda=\text{const}$, 
yields the familiar exponential collapse~\cite{biskamp_magnetic_2005}. 
Our approach differs in that we replace the arbitrary function $\Lambda(t)$ with $\vdr/L(t)$, 
so that the outflow velocity is $u_y=\partial \phi/\partial x = \vdr y/L(t)$. }).
Substituting these expressions into the ideal reduced-MHD equations~\cite{strauss_nonlinear_1976}, one obtains
\be
\label{eq_CK}
a(t)= {a_0 L_0}/(L_0 + 2\vdr t) ,\quad L(t) = L_0 + 2\vdr t, 
\ee
where $a_0\equiv a(0)$, $L_0\equiv L(0)$.
The CS formation driving timescale then becomes $\tdr=L/\vdr \sim t$ for $t\gg L_0/\vdr$.
The two main dimensionless parameters of the system are  
the Alfv\'en Mach number (quantifying the ideal-MHD CS formation drive), 
$\Mdr\equiv \vdr/V_A$, assumed to be $\leq$ 1, and the initial Lundquist number $S_0\equiv (a_0L_0)^{1/2}V_A/\eta \gg 1$.

Focusing on late times, $2\vdr t \gg L_0$ (and hence $L \gg a$),
we see that the tearing instability parameter is
$\Dprime(t) = (16/N) \, [\vdr^3/(a_0^2 L_0^2)]\, t^3$.
The transition from the FKR to the Coppi regime for the $N=1$ mode occurs when $(L/a) S_a^{-1/4}\sim 1$, i.e.,
$\ttr(1)/\tauAck\sim \Mdr^{-1}S_0^{1/9}$,
where $\tauAck \equiv (a_0 L_0)^{1/2}/V_A$.
The critical time for the $N=1$ FKR mode, $\gmaxFKR L(\tcr)/\vdr\sim 1$, is
$\tcrFKR(1)/ \tauAck \sim \Mdr^{-12/17}S_0^{3/17}$~\footnote{
As a side remark, the current sheet's aspect ratio at this time, 
$[L/a]_{\rm cr}^{\rm FKR} \sim \Mdr^{10/17} S_0^{6/17}$, can be expressed in terms of 
$S_a(\tcrFKR) \sim \Mdr^{-5/17}\, S_0^{14/17}$ as 
$[L/a]_{\rm cr}^{\rm FKR} \sim \Mdr^{5/7}\, [S_a(\tcrFKR)]^{3/7}$, which generalizes Bulanov {\it et al}'s \cite{bulanov_tearing_1979} Alfv\'enic-drive ($\Mdr=1$) result $L/a \sim S_a^{3/7}$ to the case of arbitrary drive~$\Mdr$ and shows the mutual consistency of ours and theirs approaches.
}.
The condition $\tcrFKR(1) <\ttr(1)$ for the fastest growing linear ($N=1$) mode to be in the FKR regime at this time yields a condition on the drive:
\be
\label{eq_CK-Mach-crit}
\Mdr<\Mdrc\equiv S_0^{-2/9}.
\ee
If this is not satisfied, then the $N=1$ mode and several higher-$N$ modes transition to the Coppi regime while still in the linear stage. The fastest growing mode number then increases with time as $\NmaxCoppi \sim(L/a) S_a^{-1/4}$.
As argued above, the mode that ``wins" the linear stage is the mode with~$\ttr \simeq \tcr$. 
For the CS formation model yielded by (\ref{eq_CK}), this dominant Coppi mode is 
\be
\label{eq_NmaxCoppi-CK}
\NmaxCoppi\sim \[L/a S_a^{-1/4}\]_{t=\tcrCoppi}\sim \Mdr^{9/10}S_0^{1/5}\,  ,
\ee
and its critical time and CS dimensions  at that time are  
\bea
\label{t_cr_CK-Coppi}
\tcrCoppi \simeq \ttr(\NmaxCoppi) & \sim & \Mdr^{-3/5} S_0^{1/5}\, \tauAck \, , \\
\label{a_cr_CK-Coppi}
a_{\rm cr}^{\rm Coppi}/(a_0L_0)^{1/2} & \sim &\Mdr^{-2/5}S_0^{-1/5},\\
\label{L_cr_CK-Coppi}
L_{\rm cr}^{\rm Coppi}/(a_0L_0)^{1/2} & \sim &\Mdr^{2/5}S_0^{1/5},
\eea
and hence $(L/a)_{\rm cr}^{\rm Coppi} \equiv (L/a)|_{\tcrCoppi} \sim \Mdr^{4/5}S_0^{2/5}$.
As a consistency check, we see that $\ttr(1)\ll \tcrCoppi$ if $\Mdr\gg S_0^{-2/9}$, which is required for the dominant mode to be in the Coppi regime by the end of the linear stage in the first place.
Also, it is instructive to note that $(L/a)_{\rm cr}^{\rm Coppi} \sim [S_L(\tcrCoppi)]^{1/3}\Mdr^{2/3}$ (where $S_L(t) \equiv L(t) V_A/\eta$),  which  generalizes the scaling obtained in Ref.~\cite{pucci_reconnection_2014} for $\Mdr=1$. 

In the nonlinear phase, if the condition (\ref{eq_CK-Mach-crit}) is satisfied, the $N=1$ mode continues to dominate and undergoes Rutherford evolution described by $\dot{w}_1 \simeq \eta \Delta'(N=1)$, yielding 
$w_1(t) \simeq w_1(\tcrFKR) + 4 \eta\, \vdr^3\, (t^4 - [\tcrFKR]^4)/(a_0^2 L_0^2)$, 
where $w_1(\tcrFKR) \simeq \delta_{in}^{\rm FKR}$ is the island width at the start of the Rutherford stage.
The Rutherford stage continues until $X$-point collapse and it can be checked {\it a posteriori} that, if (\ref{eq_CK-Mach-crit}) is satisfied, then the Rutherford stage lasts much longer than~$\tcrFKR(1)$. 
One can then also show that the critical island width triggering $X$-point collapse, $w_{X,1}$, 
is much greater than~$w_1(\tcrFKR)$. 
Thus, the growth of the $N=1$ FKR mode throughout most of the Rutherford stage
is described by (ignoring factors of order unity):
\be
\label{eq_w_t_CK-simple}
w_1(t) \sim  \eta\,{{\vdr^3}\over{a_0^2 L_0^2}}\, t^4   \, .
\eeq

Then, the time for this mode to reach the collapse size $w_{X,1} = [\Delta'_1(t_{X,1})]^{-1}$ 
is 
\be
t_{X,1} \sim \tauAck \Mdr^{-6/7}S_0^{1/7} \gg \tcrFKR \, ,
\ee
the critical island width is $w_{X,1} \sim (a_0L_0)^{1/2}S_0^{-3/7}\Mdr^{-3/7}$,
and the corresponding CS parameters are 
\bea
\label{a_X_CK-FKR}
a_{X,1}/(a_0L_0)^{1/2}& \sim &\Mdr^{-1/7}S_0^{-1/7},\\
\label{L_X_CK-FKR}
L_{X,1}/(a_0L_0)^{1/2}& \sim &\Mdr^{1/7}S_0^{1/7},\\
\label{asp_rat_CK-FKR}
\[L/a\]_{X,1}& \sim &\Mdr^{2/7}S_0^{2/7}.
\eea
As we have argued, the mode's growth accelerates rapidly after the collapse~\cite{loureiro_x-point_2005} and so $\tdisruptFKR \sim t_{X,1}$; thus, the above expressions yield good practical estimates for the final parameters at the moment of disruption $w = a$. 

If, on the other hand, \eq{eq_CK-Mach-crit} is not satisfied, then both linear and nonlinear stages are dominated by the Coppi mode with $\NmaxCoppi$ given by~\eq{eq_NmaxCoppi-CK}.  
This mode undergoes $X$-point collapse and then quickly leads to CS disruption essentially as soon as it becomes nonlinear, i.e., $\tdisruptCoppi\simeq t_X^{\rm Coppi} \sim \tcrCoppi$. 
Consequently, the CS thickness $a_{X}^{\rm Coppi}$ and length $L_{X}^{\rm Coppi}$ at disruption are well approximated by their values~\eqs{a_cr_CK-Coppi}{L_cr_CK-Coppi} at~$\tcrCoppi$.

It is worth noting that both \eqs{a_X_CK-FKR}{asp_rat_CK-FKR} and \eqs{a_cr_CK-Coppi}{L_cr_CK-Coppi} scale only weakly with the two key input parameters $\Mdr$ and~$S_0$, pointing to a certain universality of the FKR/Rutherford and the Coppi evolution scenarios: in each of these regimes one will find reasonably similar estimates for a wide range of $\Mdr$ and~$S_0$. 

Also note that in both the FKR and Coppi cases $a_X$ is much larger than the corresponding SP CS thickness $\delta_{\rm SP} \sim L_X S_X^{-1/2}$, where  $S_X\equiv L_X V_A/\eta$ is the Lundquist number at the time of $X$-point collapse.  The ensuing CS disruption implies, therefore, that a global-scale SP layer is never formed, as we anticipated.


As an application, let us consider typical parameters for solar flares:
$a_0= L_0 = 10^4\,\text{km}$, $n_e = 10^{10} \, {\rm cm^{-3}}$, $B_0=100$~G, resulting in $V_A\simeq 2000\,\text{km/s}$, $\tau_{A0} \simeq 5\, {\rm s}$, and $S_0 \approx 3\times 10^{13}$.
\eq{eq_CK-Mach-crit} yields roughly $\Mdrc\approx 0.001$, corresponding to 
$v_{\rm dr,c} \simeq 2 \, {\rm km/s}$, comparable to typical photospheric velocities.
Since in the real corona a broad range of drives is likely to be present, let us contemplate both the FKR and Coppi cases by considering $\Mdr=\Mdrc= 10^{-3}$ (FKR, $N\sim 1$) and $\Mdr=0.05$ (Coppi; 
in this case $v_{\rm dr} \approx 100$ km/s, as may arise due to ideal-MHD instabilities or a loss of equilibrium driving a coronal mass ejection). 
We obtain: 
(i) $\Mdr=10^{-3} \rightarrow a_{\rm disrupt}^{\rm FKR}\approx 300\,\text{km},~
L_{\rm disrupt}^{\rm FKR}\approx 3\times 10^5\,\text{km}$, $\tdisruptFKR \approx 40\,\text{hours}$; 
(ii) $\Mdr=0.05\rightarrow a_{\rm disrupt}^{\rm Coppi}\approx 70\,\text{km},~
L_{\rm disrupt}^{\rm Coppi}\approx 1.5\times 10^6\,\text{km},~\tdisruptCoppi\approx 4\,\text{hours}$, 
and $N\approx 30$.
These are reasonable numbers (see, e.g., \cite{shibata_solar_2011, bemporad_spectroscopic_2008, ciaravella_current_2008, barta_spontaneous_2011}), especially in light of the crudeness of the CS formation model considered here.
In particular, both $\tdisruptFKR$ and $\tdisruptCoppi$ are consistent with observed pre-flare energy-buildup times.
In addition, note that in both cases, the smallest meaningful length scale in our problem, 
$\deltain (t_{\rm cr})\sim 100-300\,\text{m}$, remains much larger than the ion kinetic scales:  
the skin depth, $c/\omega_{pi}\approx 2\,\text{m}$, and the Larmor radius, 
$\rho_i\approx 0.1\,\text{m}$.  
This validates our usage of the resistive MHD description for reconnection onset in the solar corona in this example.


\paragraph{Conclusions.}
In this study we have developed a general conceptual framework connecting two important and related phenomena that have hitherto been considered separately: large-scale ideal-MHD processes leading to thin current sheet formation and magnetic energy accumulation, and the onset of fast energy release through reconnection. 
In our picture, the immediate outcome of this sequence of events is the disruption (and thus replacement) of the forming current sheet by a chain of primary magnetic islands generated by the tearing instability.  
Our study is substantially different from, and more fundamental than, previous related work on the tearing instability of reconnecting current sheets~\cite{bulanov_tearing_1979, pucci_reconnection_2014}, which has focused exclusively on the linear evolution of a time-independent current sheet and has not considered if and how the FKR regime transitions into the Coppi regime. 
In contrast, we have considered a time-evolving current sheet at an arbitrary formation rate, computed the pertinent timescales related to various tearing modes, and analyzed the order in which these various processes happen during both the linear and nonlinear evolution.
Our analysis has allowed us to predict for the first time the moment at which the current sheet is disrupted (the reconnection onset), the number of primary magnetic islands that disrupt it, and the final current sheet properties at the time of disruption, and elucidate their dependence on the Lundquist number and the current-sheet formation rate.  
In particular, our analysis has revealed that two distinct regimes are possible: 
the FKR/Rutherford regime, in which the sheet is disrupted by only one or two islands; 
and the Coppi regime, where, instead, it is disrupted by a large number of islands.  
Both scenarios are relevant to experimental, astrophysical, and space systems, including solar flares, where they yield reasonable estimates for flare onset~\cite{shibata_solar_2011}.

Although here we have restricted ourselves to the resistive MHD description, the conceptual framework outlined in this Letter is completely general and can be extended to collisionless plasmas, provided that the linear and nonlinear regimes of the tearing instability are understood in the particular collisionless formulation that one chooses to adopt; this is indeed necessary for addressing the onset problem in two prominent contexts: sawtooth crashes in tokamaks~\cite{hastie_sawtooth_1997} and reconnection in the Earth's magnetotail~\cite{bhattacharjee_impulsive_2004}.


\paragraph{Acknowledgments.}
NFL was partially supported by Funda\c{c}\~ao para a Ci\^{e}ncia e a Tecnologia 
through grants IF/00530/2013, PTDC/FIS/118187/2010 and Pest-OE/SADG/LA0010/2011.


\bibliography{UL-2014_resub}

\end{document}